\documentstyle[psfig,12pt]{article}
\def\ltsima{$\; \buildrel < \over \sim \;$}
\def\simlt{\lower.5ex\hbox{\ltsima}}
\def\gtsima{$\; \buildrel > \over \sim \;$}
\def\simgt{\lower.5ex\hbox{\gtsima}}

\begin{document}
\begin{center}
{\large \bf Cosmic Ray Anisotropy Analysis with a Full-Sky Observatory\\}

\vspace{1cm}
{\large \it Paul Sommers\\}
\vspace{3mm}
{\it High Energy Astrophysics Institute\\
University of Utah Physics Department\\
115 S 1400 E, Room 201\\
Salt Lake City, UT 84112-0830}

\vspace{1cm}
{\bf Abstract}
\end{center}

\noindent A cosmic ray observatory with full-sky coverage can
exploit standard anisotropy analysis methods that do not work if
part of the celestial sphere is never seen.  In particular, the
distribution of arrival directions can be fully characterized by
a list of spherical harmonic coefficients.  The dipole vector and
quadrupole tensor are of special interest, but the full set of
harmonic coefficients constitutes the anisotropy fingerprint that
may be needed to reveal the identity of the cosmic ray sources.
The angular power spectrum is a coordinate-independent synopsis
of that fingerprint.  The true cosmic ray anisotropy can be
measured despite non-uniformity in celestial exposure, provided
the observatory is not blind to any region of the sky.  This
paper examines quantitatively how the accuracy of anisotropy
measurement depends on the number of arrival directions in a data
set.

\vspace{1cm}
\section{Introduction}

The origin of the highest energy cosmic rays is a problem that
has persisted for 4 decades since the pioneering measurements at
Volcano Ranch \cite{L-S,Linsley}.  There is some consensus that,
above the spectrum's ankle at about $10^{18.5}$ eV, they
originate outside the disk of the Galaxy.  For particles of such
high magnetic rigidity, sources in the Galaxy's disk would
presumably cause an obvious anisotropy in arrival directions that
is not observed.  Evidence for a composition changing to lighter
particles at the ankle \cite{FE} strengthens this argument
(particles having even greater rigidity because of lesser charge)
and supports the view that cosmic rays with energies above the
ankle are of extragalactic origin.  The sources of those
particles remain to be identified.

The observations of cosmic rays with energies above the expected
GZK cutoff \cite{Greisen,Z-K} should be a powerful clue to the
nature of the sources.  The Fly's Eye \cite{Big1} and AGASA
\cite{Big2} measured air showers with energies well above the GZK
threshold.  Recent reports \cite{AGASA,Jui,Watson} suggest that the
spectrum might continue without a strong GZK effect.  These
super-GZK results have posed several related, but
distinguishable, puzzles:

\begin{enumerate}
\item{How are particles produced with such prodigious energy?}
\item{Why do the arrival directions of those particles not point
back to recognizable sources in our local part of the universe?}
\item{Why is the intensity of particles above $6\times 10^{19}$
eV not more strongly suppressed?}
\end{enumerate}

New attempts to improve the observational data include the
recently commissioned High Resolution Fly's Eye (HiRes)
\cite{HiRes}, the Pierre Auger Observatory \cite{Auger}, the
proposed Telescope Array Project \cite{TA}, and Airwatch/OWL
\cite{OWL} plans for future space-based detectors of atmospheric
air showers.  The Auger Observatory and the potential space-based
detectors will have exposure to the entire sky, which will open
new possibilities for anisotropy analysis.  These methods will be
explored here in the context of the better controlled exposure of
the Auger surface arrays.

Hillas \cite{Hillas1} pointed out years ago that there are few astrophysical
sites that can produce a large enough ``electrical potential''
$\beta B L$ to accelerate even highly charged nuclei to $10^{20}$
eV.  (Here $\beta=v/c$ is the relative velocity of moving media
with magnetic field strength $B$ and size $L$.  In the case of
statistical acceleration, including Fermi shock acceleration, the
same product $\beta B L$ governs the maximum particle rigidity
even though particles do not pass through a monotonic change in
electrical potential.)  Puzzle \#1 is exacerbated in many
contexts by synchrotron radiation and/or pion photoproduction.

Puzzle \#2 can be resolved by invoking stronger-than-expected
extragalactic magnetic fields, but that does not readily simplify
puzzle \#3.  The expected suppression of particles above the GZK
cutoff is based on travel time (or total distance traveled)
rather than the straight-line distance to the sources.  Particles
below the GZK threshold have been accumulating over billions of
years, whereas the mean age of particles well above the threshold
cannot be greater than tens of millions of years \cite{Elbert}.  

One possible inference from the lack of an observable GZK
spectral break is that sub-GZK particles may not be much older
than 30 million years either, in which case the GZK effect would
not significantly suppress the cosmic rays above the threshold
relative to those below.  It is not reasonable to suppose that
the sources of high energy cosmic rays turned on so recently, so
the young age of sub-GZK cosmic rays would require an
intergalactic mechanism for dissipating their energy.  Known
mechanisms (e.g. nuclear interactions, synchrotron radiation,
$e^{\pm}$ production, pion photoproduction via infrared and
visible background photons, etc.) do not rob energy from sub-GZK
particles rapidly enough.  Perhaps high energy cosmic rays are
attenuated through interactions with the unknown dark matter of
the universe \cite{Sommers}.

More conventional approaches to puzzle \#3 are to defeat the GZK
cutoff with a very hard extragalactic spectrum (e.g. from
topological defect annihilation \cite{Schramm}) or to evade it by
invoking neutrinos \cite{Yoshida,Weiler} or non-standard
particles \cite{Farrar1,Farrar2} that are immune to the microwave
background radiation.  Others \cite{Berezinsky,Hillas2} conclude
that the sources must be localized to the Galaxy, but distributed
in a halo large enough that galactic anisotropy has not become
obvious.

The cumulative cosmic ray observations at this time are not
sufficient to sort out the possibilities.  AGASA and HiRes are
currently building up the world's total exposure at the highest
energies.  With better statistics and better measurements, the
observations could soon lead to a breakthrough that identifies
the sources of the highest energy cosmic rays.  This same hope
has been expressed for decades through the course of numerous
experiments, however, and the puzzles have only become deeper
mysteries.  The answers may not come easily, and we should
prepare the best possible analyses of the energy spectrum,
particle mass distribution, and arrival directions.  

Careful determinations of the energy spectrum and mass
composition can be used to weed out classes of theories, but
these tools are not likely to yield a clear signature for picking out a
unique theory.  A positive identification of the cosmic ray
sources requires seeing their fingerprint in the sky.  This may
come in the form of arrival direction clusters
\cite{Clusters,Chi} that identify discrete sources, or it may
come as a large-scale celestial pattern that characterizes a
particular class of potential sources.  In the worst case, we
might discover that the arrival directions are isotropic and the
sources still elude positive identification.  In that case,
observers must strive for the best possible upper limit on
anisotropy.

The Auger Project's surface arrays will provide the best search
for anisotropy fingerprints.  Their combined exposure function on
the celestial sphere will be unambiguous because they operate
continuously and are not sensitive to atmospheric variability.
Continuous operation means the celestial exposure function is
uniform in right ascension.  By having observatories
appropriately located in both the southern and northern
hemispheres, the exposure does not vary strongly with declination
either.  

The methods described in this paper are applicable to any
observatory with full-sky coverage.  They are not limited to the
Auger Project, although the specific Auger site locations are
used in the example simulations that are reported here.  

Coverage of the full sky could be achieved piecemeal by combining
results from different experiments.  There is serious risk of
spurious results from such meta-analyses, however, unless the
exposures, energy resolutions, and detector systematics are
perfectly understood and correctly incorporated in the analysis.
The reliable approach is to use identical detectors in both
hemispheres or the same (orbiting) detector for both hemispheres.

This paper seeks to evaluate the sensitivity of a full-sky
observatory to large-scale anisotropy patterns and how that
sensitivity depends on the number of arrival directions in a data
set.  Large-angle fingerprinting will be needed if there are many
contributing sources or if the flux from each single source is
diffused over a large solid angle due to magnetic deflection of
the charged cosmic rays.  If, instead, there are point sources to
be detected, then the advantage of a full-sky observatory is in
mapping the entire celestial sphere with comparable sensitivity
in all regions.

Full-sky coverage is crucial for large-scale anisotropy analysis.
It makes it possible to do integrals over the sky, so the
powerful tools of multipole moments and angular power spectra are
available.  With full sky coverage, cosmic ray anisotropy
analysis will be similar to gamma ray burst anisotropy analysis.
The numbers of events will be comparable, the direction error
boxes will be comparable, the exposure non-uniformities will be
comparable, and in both cases events come from all parts of the
sky.  All of the techniques that were employed to search for
anisotropy in the BATSE data \cite{Batse1,Batse2} can be applied
to a full-sky cosmic ray data set.

With a cosmic ray detector in only one hemisphere, there is a
solid angle hole in the sky where the detector has zero exposure
despite the Earth's rotation.  A zero-exposure hole makes it
impossible to do integrals over the whole celestial sphere.  No
matter how many events the detector collects overall, it will
never determine any multipole moment.  A single-hemisphere
detector can test hypotheses like, ``Does the observed
distribution match better what would be accepted from the
clustering of radio galaxies toward the supergalactic plane or
what would be accepted from an isotropic distribution?''  It can
also make qualified measurements like, ``Assuming the anisotropy
is a perfect quadrupole with axial symmetry, fit for the axis
orientation that best explains the observed celestial
distribution.''

The role of an observatory, however, should be to map the sky and
make results available in a form which is readily usable without
knowledge of the detector properties and which is
independent of any theoretical hypothesis.  Low-order multipole
tensors (or spherical harmonic coefficients) can summarize the
large-scale information.  The angular power spectrum reveals if
there is clumpiness on smaller scales.  These results can be
tabulated so that theorists can test arbitrary models
quantitatively without privileged access to the data.  With
approximately uniform exposure, even eyeball inspection of
arrival direction scatter plots can show large-scale patterns
that are hidden when steep exposure gradients dominate the
scatter plots.  

While the role of an observatory should be to map the sky and
determine the patterns without preconceived expectations, it is
nevertheless worthwhile to consider what might be learned by
measuring the low order multipoles or the angular power
spectrum.  

\vspace{.3cm}
{\it Monopole.}  There is no information about anisotropy
patterns in the monopole scalar by itself.  It is simply the sky
integral of the cosmic ray intensity.  That is information
already present in the energy spectrum.  A pure monopole
intensity distribution is equivalent to isotropy.  The strength
of other multipoles relative to the monopole is a measure of
anisotropy.  

\vspace{.3cm} {\it Dipole.}  A pure dipole distribution is not
possible because the cosmic ray intensity cannot be negative in
half of the sky.  A ``pure dipole deviation from isotropy'' means
a superposition of monopole and dipole, with the intensity
everywhere $\geq 0$.

A predominantly dipole deviation from isotropy might be expected
if the sources are distributed in a halo around our Galaxy, as
has been suggested \cite{Berezinsky,Hillas2}.  In this case,
there is a definite prediction that the dipole vector should
point toward the galactic center.

An approximate dipole deviation from isotropy could be caused by
a single strong source if magnetic diffusion or dispersion
distributes those arrival directions over much of the sky.  In
general, a single source would produce higher-order moments as
well.

A dipole moment is measurable in the microwave radiation due to
Earth's motion relative to the universal rest frame \cite{COBE}.  If we
are moving relative to the cosmic ray rest frame, a dipole moment
should exist also in the cosmic ray intensity (the
Compton-Getting effect).  At lower energies, this may occur if
the sun and Earth are moving relative to the galactic magnetic
field or if the cosmic rays are not at rest with respect to the
galactic field.  For extragalactic cosmic rays, a Compton-Getting
dipole is expected if the Galaxy is moving relative to the
intergalactic field or if the cosmic rays themselves are
streaming in intergalactic space.  In any case, the expected
velocities would be small ($v/c \simlt 10^{-3}$), and the Compton
Getting anisotropy \cite{Longair}
($\frac{I_{max}-I_{min}}{I_{max}+I_{min}}=(\gamma+2)(v/c)$) should
be $\simlt 0.005$.  (Here $\gamma$ is the differential spectral
index, which is roughly 3.)  An anisotropy of one-half percent
would require high statistics for detection (see section 3).

A larger dipole anisotropy might be produced by a cosmic ray
density gradient.  If the magnetic field is disorganized, the
gradient produces streaming by diffusion and the Compton-Getting
dipole vector is parallel to the density gradient.  If there is a
regular magnetic field, however, the expected dipole vector $\vec{D}$ can
be perpendicular to both the gradient and the field direction, 
$\vec{D} \propto \vec{\nabla} \rho \times \vec{B}$.
The direction of strongest intensity corresponds to the arrival
direction of particles whose orbit centers are located in the
direction of increased density.

\vspace{.3cm}
{\it Quadrupole.}  An equatorial excess in galactic coordinates
or supergalactic coordinates would show up as a prominent
quadrupole moment.  A measurable quadrupole is expected in many
scenarios of cosmic ray origins, and is perhaps to be regarded as
the most likely result of a sensitive anisotropy search.  

In general, a quadrupole tensor is characterized by 3 relative
eigenvalues with associated orthogonal eigenvectors.  In the case
of axial symmetry, there is a single non-degenerate eigenvector
that gives the symmetry axis.  An axisymmetric ``prolate''
distribution would be hot spots at antipodal points of the sky,
whereas an ``oblate'' distribution has the excess concentrated
toward the equator that is perpendicular to the symmetry axis.

The axis of an oblate quadrupole distribution might differ from
the galactic axis or the supergalactic axis if we are embedded in
a magnetic field that systematically rotates the arrival
directions.

\vspace{.3cm}
{\it The angular power spectrum.}  Spherical harmonic
coefficients for a function on a sphere are the analogue of
Fourier coefficients for a function on a plane.  Variations on an
angular scale of $\theta$ radians contribute amplitude in the
$\ell=1/\theta$ modes just as variations of a plane function on a
distance scale of $\lambda$ contribute amplitude to the Fourier
coefficients with $k=2\pi/\lambda$.

For cosmic ray anisotropy, we might look for power in modes from
$\ell = 1$ (dipole) out to $\ell \sim 60$, higher order modes
being irrelevant because the detector will smear out any true
variations on scales that are smaller than its angular
resolution.  For charged cosmic rays, magnetic dispersion will
presumably smear out any point source more than the detector's
resolution function.  Even at the highest observed particle
energies, there is unlikely to be any structure in the pattern of
arrival directions over angles smaller than $3^{\circ}$.  The
interesting angular power spectrum is therefore probably limited
to $\ell \simlt 20$.

\section{Exposure}

For a cosmic ray observatory, exposure is a function on the
celestial sphere.  Measured in units $km^2\cdot yr$, it gives the
observatory's time-integrated effective collecting area for a
flux from each sky position.  In this paper, the {\it relative}
exposure $\omega$ is usually the function of interest.  That will be a
dimensionless function on the sphere whose maximum value is 1.
In other words, $\omega$ at any point of the sky
is a fraction between 0 and 1 given by the exposure at that point
divided by the largest exposure on the sky.

In other contexts, the term ``exposure'' refers to the total
exposure integrated over the celestial sphere.  It then has units
$km^2\cdot sr\cdot yr$.  For example, in determining the cosmic ray
energy spectrum, one divides the number of cosmic rays
observed in each energy bin by the total exposure for that
energy.  (In general, an observatory's exposure is energy
dependent.)  If there were evidence that the energy spectrum were
not uniform over the sky, then we would need to use the exposure's
dependence on celestial position to map the spectrum over the
sky.  

Since the spectrum is defined by the number of observed events
divided by total exposure, one can use the measured spectrum to
get the expected number of cosmic rays for any given total exposure.
In the case of the Auger surface arrays, the continuous
acceptance is approximately $14,000\ km^2sr$, independent of
energy above $10^{19}$ eV.  After operating for 5 years, they
will have a total exposure of $70,000\ km^2\cdot sr\cdot yr$.
The integral cosmic ray intensity above $10^{19}$ eV is approximately
$0.5/(km^2\cdot sr\cdot yr)$, and it falls roughly like $E^{-2}$
(perhaps less rapidly, but the energy dependence is not well
determined above $6\times 10^{19}$ eV.)  Using this simple
$E^{-2}$ dependence gives the following estimates for Auger
cosmic ray counts after 5 years:

\begin{itemize}
\item{35,000 above $10^{19}$ eV (believed to be mostly extragalactic)}
\item{2200 above $4\times 10^{19}$ eV (compared to 47 in the
AGASA cluster analysis)}
\item{350 above $10^{20}$ eV (above the GZK threshold region)}
\item{35 above $3.2\times 10^{20}$ eV (highest energy measured so
far)}
\end{itemize}

How any number of detected cosmic rays are distributed on the sky depends on
both the true celestial anisotropy and the observatory's relative
exposure $\omega$.  

The relative exposure can be calculated as follows for a detector at
a single site with continuous operation.  Full-time operation
means that there is no exposure variation in sidereal time and
therefore constant exposure in right ascension.  Suppose the
detector is at latitude $a_0$ and that 
it is fully efficient for particles arriving with zenith angles
$\theta$ less than some maximum value $\theta_m$.  (Full
efficiency means the zenith angle acceptance depends on zenith
angle only due to the
reduction in the perpendicular area given by $cos(\theta)$.)
This results in the following dependence on declination $\delta$:
$$\omega(\delta)\ \propto\ cos(a_0) cos(\delta) sin(\alpha_m)+
        \alpha_m sin(a_0) sin(\delta),$$
where $\alpha_m$ is given by 
$$ \alpha_m = \left\{ \begin{array}{ll} 
\ 0 & \mbox{if $\xi > 1$}\\
\ \pi & \mbox{if $\xi < -1$}\\
\ cos^{-1}(\xi) & \mbox{otherwise}
 \end{array} \right.  $$
and
$$\xi \equiv
\frac{cos(\theta_m)-sin(a_0)sin(\delta)}{cos(a_0)cos(\delta)}.$$

\begin{figure}
\psfig{figure=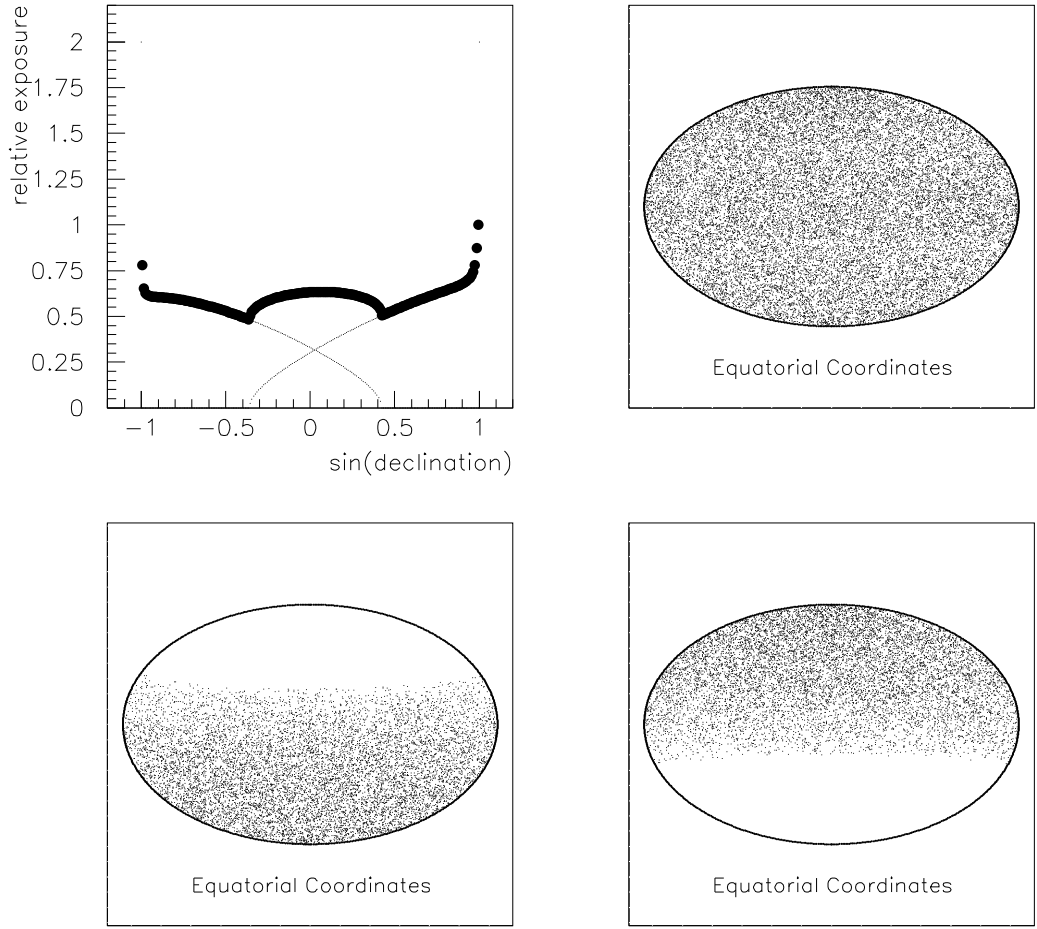,width=6.0in}
\vspace{1cm}
\caption{The upper left plot shows the declination dependence of
the Auger Observatory's relative exposure.  The southern and
northern sites are indicated separately by dots.  The combined
exposure function is marked with filled circles.  The scatter
plot in the upper right results from sampling an isotropic
distribution and applying this Auger acceptance.  There are
20,000 directions plotted, 10,000 from the southern site (shown
separately in the lower left plot) and 10,000 from the northern
site (lower right).}
\end{figure}

The upper left plot of figure 1 shows the resulting declination
dependence for a site at $a_0=-35^{\circ}$ and another site at
$a_0=+39^{\circ}$, which are the latitudes for the two Auger
observatories.  The detectors are assumed to be fully efficient
out to $\theta_m=60^{\circ}$, and no arrival directions are
counted from larger zenith angles.  The combined exposure is also
shown.  The maximum is at the north pole direction, which is
always detectable at the northern site, although the effective
detector area is reduced by $cos(51^{\circ})$ for flux arriving
from the north pole direction.

The lower plots in figure 1 are scatter plots of the accepted
cosmic rays for each site, where directions have been sampled
from an isotropic distribution but accepted according to each
detector's exposure.  (A sampled cosmic ray direction is accepted
if a randomly sampled number between 0 and 1 is less than the
relative exposure $\omega$ for that direction.)  There are 10,000
accepted arrival directions in each of those two plots.  Shown in
the upper right plot is the superposition of all 20,000 events
from the combined observatory.  The combined distribution is not
uniform, but has the modest declination dependence indicated in
the upper left plot.

\section{Dipole sensitivity}

The objective here is to study the sensitivity of a full-sky
observatory to a dipole deviation from isotropy.  How well can
the dipole be measured?  How does that accuracy depend on the
number of arrival directions in the data set?  How does it depend
on the amplitude of the dipole anisotropy?

For a dipole deviation from isotropy, the cosmic ray intensity
varies over the sky as
$$I(\vec{u})= \frac{N}{4\pi}(1+\alpha \vec{D}\cdot\vec{u}).$$
Here $\vec{u}$ is a unit vector defining the celestial direction,
$\frac{N}{4\pi}$ is the average intensity, $\vec{D}$ is the
dipole direction unit vector, and $\alpha$ is its (non-negative)
amplitude.  In order for the cosmic ray intensity to be nowhere
negative, $\alpha$ must lie in the range $0 \leq \alpha \leq 1$.
The amplitude $\alpha$ gives the customary measure of anisotropy
amplitude: $\alpha=(I_{max}-I_{min})/(I_{max}+I_{min})$.

The dipole can be recovered from the celestial intensity function
by
$$\alpha \vec{D} = \frac{3}{N}\ \int I(\vec{u})\ \vec{u}\
d\Omega.$$ In our case, the observed intensity function consists
of $N$ discrete arrival directions, each associated with a
relative exposure $\omega_i$.  The components of the dipole vector
are then estimated by
$$\alpha D_a = \frac{3}{\cal N}\ \sum_{i=1}^{N}\
\frac{1}{\omega_i}\ u^{(i)}_a, $$ where $u^{(i)}_a$
denotes a component of the $i$th vector, and $\cal N$ is the
simple sum of the $N$ weights $\frac{1}{\omega_i}$.  (These
dipole components are linear combinations of the three spherical
harmonic coefficients with $\ell=1$.)

To test this method's sensitivity to a dipole of amplitude
$\alpha$ when there are N directions in the data set, one can
produce an ensemble of artificial data sets of this type (with
random dipole directions $\vec{D}$).  For each data set, use the
above formula to estimate the dipole vector, and record the
difference of the estimated $\alpha$ from the input alpha and
also the angle between the estimated direction and the input
dipole direction.  These error distributions describe the
measurement accuracy.  The RMS deviation from the true $\alpha$
is a single number to characterize the amplitude measurement
accuracy, and the average space angle error summarizes the
accuracy of determining the dipole direction.

This procedure can be repeated for different values of $N$ and
different values of $\alpha$.  For any pair ($N,\alpha$) the
ensemble of simulation data sets yields the amplitude resolution
and direction resolution as above.

To generate an individual simulation data set, one samples $N$
arrival directions on the celestial sphere.  First, a direction
is sampled from the assumed celestial distribution with dipole
deviation from isotropy.  Then the detector acceptance is applied
by rejecting the sampled direction if a random number is greater
than the relative exposure $\omega$ for that direction.  This
continues until the data set has $N$ arrival directions.  The
data set then reflects both the presumed celestial anisotropy and
the detector's non-uniform exposure.

These methods yield the results summarized in figure 2.  The
number of arrival directions was increased by factors of two: $N$
= 250, 500, 1000, 2000, 4000, 8000, 16000, 32000.  For each $N$,
amplitudes were studied at $\alpha$ = 0.1, 0.2,..., 1.0.

The upper left plot of figure 2 shows that the amplitude is
determined to an accuracy of about 0.1 with 250 directions,
improving to approximately 0.01 with 32,000 directions.

\begin{figure}
\psfig{figure=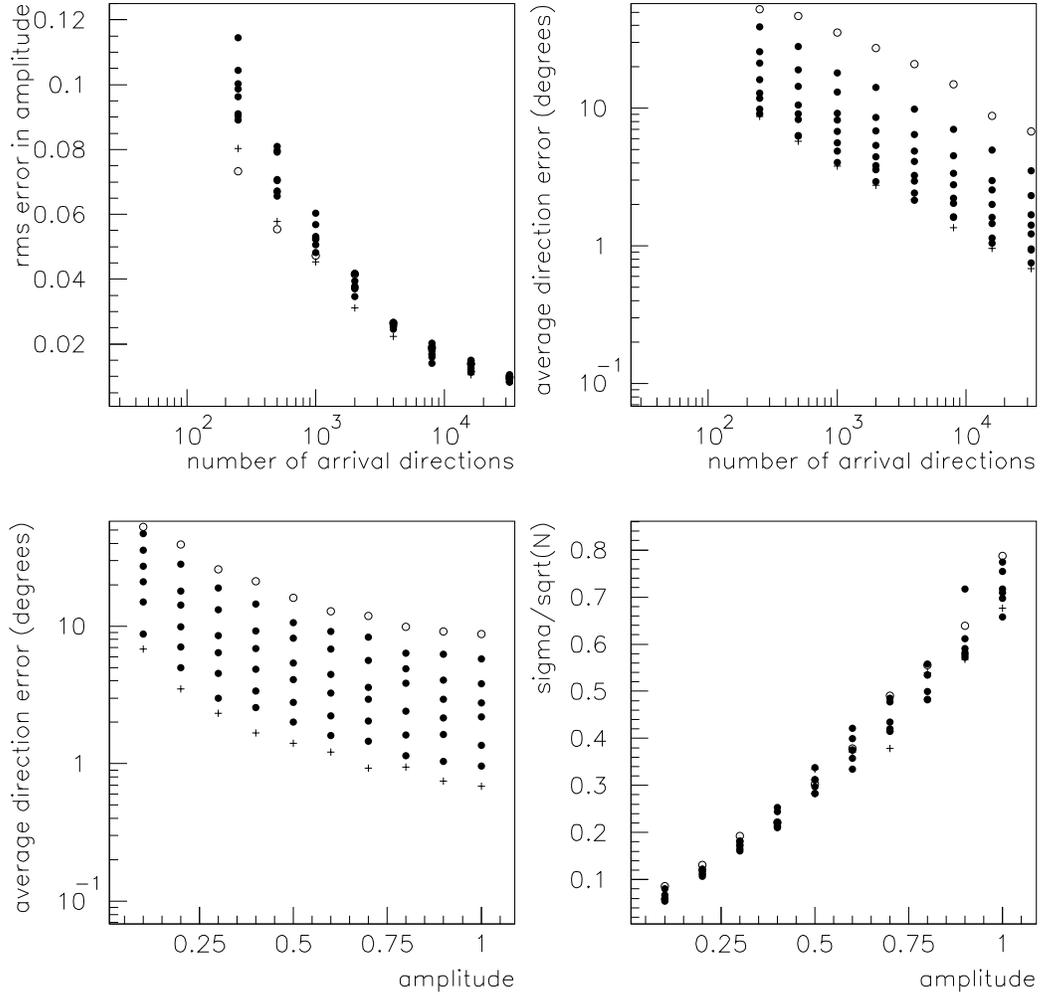,width=6.0in}
\vspace{1cm}
\caption{Four plots that indicate how the sensitivity to dipole
anisotropy depends on the number of arrival directions $N$ in the
data set and the anisotropy amplitude $\alpha$.  In the upper
plots, the abscissa is the number of directions $N$.  The left plot
shows the RMS error in estimating the amplitude: the right plot
shows the mean error in the measured dipole direction.  Each
column of points represent different amplitudes from 0.1 to 1.0.
The open circle is $\alpha=0.1$ and the + mark is $\alpha=1.0.$
In the lower figures, the abscissa is amplitude $\alpha$ and the
different points above each value represent different values of
$N$ (increasing by factors of 2).  The open circle is $N=250$,
and the + is $N=32,000$.  Multiplying the ordinate of a point in
the lower right figure by {$\protect\sqrt{N} $} gives the number of sigmas
deviation from isotropy.  Each point is derived from an ensemble of 100 
data sets.}
\end{figure}

The upper right plot shows the dipole direction resolution as a
function of the number of arrival directions.  The mean error is
less than 10 degrees for all cases (250 directions or more) if
the amplitude is nearly 1, and it is less than 10 degrees
regardless of the amplitude if the number of directions is 16,000
or more.

The lower left plot shows that the mean dipole direction error
decreases as the dipole amplitude increases.  A strong amplitude
yields a good direction determination even for a small number of
directions in the data set.  For example, with only 250 cosmic
ray arrival directions, the dipole direction is determined to
better than 20 degrees if the amplitude exceeds 0.5.  To get that
same resolution with $\alpha$=0.1, you need a data set with more
than 4000 directions.

For a fixed number of arrival directions, the RMS error in the
amplitude has little dependence on the amplitude.  That is to
say, you can distinguish amplitudes 0.85 and 0.90 as well as you
can distinguish 0.10 and 0.15.  For the purpose of detecting an
anisotropy (as opposed to measuring it), the relevant quantity is
the amplitude divided by the RMS error, which is the number of
sigmas deviation from isotropy.  That quantity increases with the
amplitude $\alpha$.  It can also be expected to increase in
proportion to $\sqrt{N}$ as the number of arrival directions
increases.  The lower right plot shows that
$$\frac{\alpha}{\Delta \alpha} \approx 0.65\ \alpha \sqrt{N}.$$
For $N$=250, for example, the deviation from isotropy increases
from 1-sigma to 10-sigma as $\alpha$ increases from 0.1 to 1.0.
For $N$=8000, the range is from 6-sigma to 60-sigma.  Etc.

To achieve a 5-sigma detection of a Compton-Getting anisotropy
amplitude of 0.005 would require 2.4 million arrival directions.
An anisotropy amplitude of 0.2 from a galactic halo distribution
of sources, however, could be detected at the 5-sigma level with
1500 arrival directions.

\section{Quadrupole sensitivity}

A quadruople deviation from isotropy is characterized by an
intensity function on the celestial sphere given by
$$I(\vec{u}) = {\bf Q}(\vec{u},\vec{u}),$$ where $\vec{u}$ is an
arbitrary direction unit vector and ${\bf Q}$ is a symmetric 2nd
order tensor.  Its trace gives the monopole moment.  Its other 5
independent components in any coordinate basis are determined
from the $\ell=2$ spherical harmonic coefficients $a_{2 m}$.

Denoting by $\xi_i$ the three eigenvalues of ${\bf Q}$ and the
three (unit) eigenvectors by $\vec{q_i}$, the intensity function has
the form
$$I(\vec{u}) = \xi_1(\vec{q_1}\cdot \vec{u})^2 +
\xi_2(\vec{q_2}\cdot \vec{u})^2 + \xi_3(\vec{q_3}\cdot
\vec{u})^2.$$ To keep the number of studied variables manageable,
consideration will be limited to axisymmetric oblate intensity
functions, as might be expected from sources in the galactic disk
or near the supergalactic plane.  Let $\vec{q}$ denote the
eigenvector that defines the symmetry axis, and let $\xi$ be the
ratio of its eigenvalue to those in the symmetry plane, so the
intensity function on the sphere is of the form
$$I(\vec{u}) \propto \xi\ (\vec{q} \cdot \vec{u})^2 +
(\vec{u}_{\bot} \cdot \vec{u}_{\bot}),$$
where $\vec{u}_{\bot} \equiv \vec{u}-(\vec{u}\cdot
\vec{q})\vec{q}$ is the part of $\vec{u}$ perpendicular to the
symmetry axis, and $0\leq \xi \leq 1$.  The anisotropy amplitude $\alpha$
is related to $\xi$ by
$$\alpha \equiv
\frac{I_{max}-I_{min}}{I_{max}+I_{min}} = \frac{1-\xi}{1+\xi}
\ \ \ \ \Leftrightarrow\ \ \ \ \xi=\frac{1-\alpha}{1+\alpha}.$$

The objective here is to test how accurately the anisotropy
amplitude $\alpha$ and the symmetry axis direction $\vec{q}$ can
be determined from a data set of arrival directions.  How does
the accuracy depend on the number of directions $N$ and on the
amplitude $\alpha$?

The method of investigation is the same as for the dipole
sensitivity study.  For each pair ($N$,$\alpha$), an ensemble of
simulation data sets are produced, each with a randomly chosen
direction for its symmetry axis.  For each data set, the arrival
directions are sampled from the relative intensity function (with
quadrupole anisotropy), and each direction is accepted with
probability equal to the relative exposure $\omega$ evaluated at
that direction.

The anisotropy amplitude $\alpha$ and symmetry axis direction
$\vec{q}$ are estimated for each simulation data set.  The tensor
with components
$$S_{ab} \equiv \int\ I\ u_a u_b\ d\Omega$$ ($u_a$ denoting a
component of $\vec{u}$) has the same eigenvectors as $Q_{ab}$ and
the components $S_{ab}$ can be estimated by
$$S_{ab} = \frac{1}{\cal N}\ \sum_{i=1}^N\ \frac{1}{\omega_i}\
u^{(i)}_a u^{(i)}_b.$$  The eigenvectors and eigenvalues of this
symmetric matrix are then found.  The symmetry axis is taken to
be defined by the eigenvector with the smallest eigenvalue.  Let
$\Delta$ be that smallest eigenvalue subtracted from the average
of the other two (which should be equal, corresponding to
directions in the symmetry plane).  The eigenvalue $\xi$ of the
intensity tensor ${\bf Q}$ is given in terms of $\Delta$ by
$$\xi=\frac{2-10\Delta}{2+5\Delta}.$$ Then the anisotropy
amplitude $\alpha$ is gotten by $\alpha = \frac{1-\xi}{1+\xi}$.

Results for ensembles with different ($N,\alpha$) values are
presented in figure 3 in complete analogy with the dipole results
presented in figure 2.  The RMS error in the amplitude and the
average space-angle error in the direction of the symmetry axis
both decrease as $N$ increases.  The symmetry axis is also seen
to be better determined as the anisotropy amplitude increases for
any fixed number of arrival directions $N$.  The sensitivity for
detecting anisotropy, as shown in the lower right plot, is given
by $$\frac{\alpha}{\Delta \alpha} \approx 0.45\ \alpha \sqrt{N}.$$ With
the definition of anisotropy amplitude $\alpha \equiv
\frac{I_{max}-I{min}}{I_{max}+I_{min}}$, detecting a quadrupole
anisotropy requires more data than for a dipole anisotropy of the
same amplitude.  Twice as many cosmic ray arrival directions
($(.65/.45)^2 \approx 2$) are needed for the same resolution.

\begin{figure}
\psfig{figure=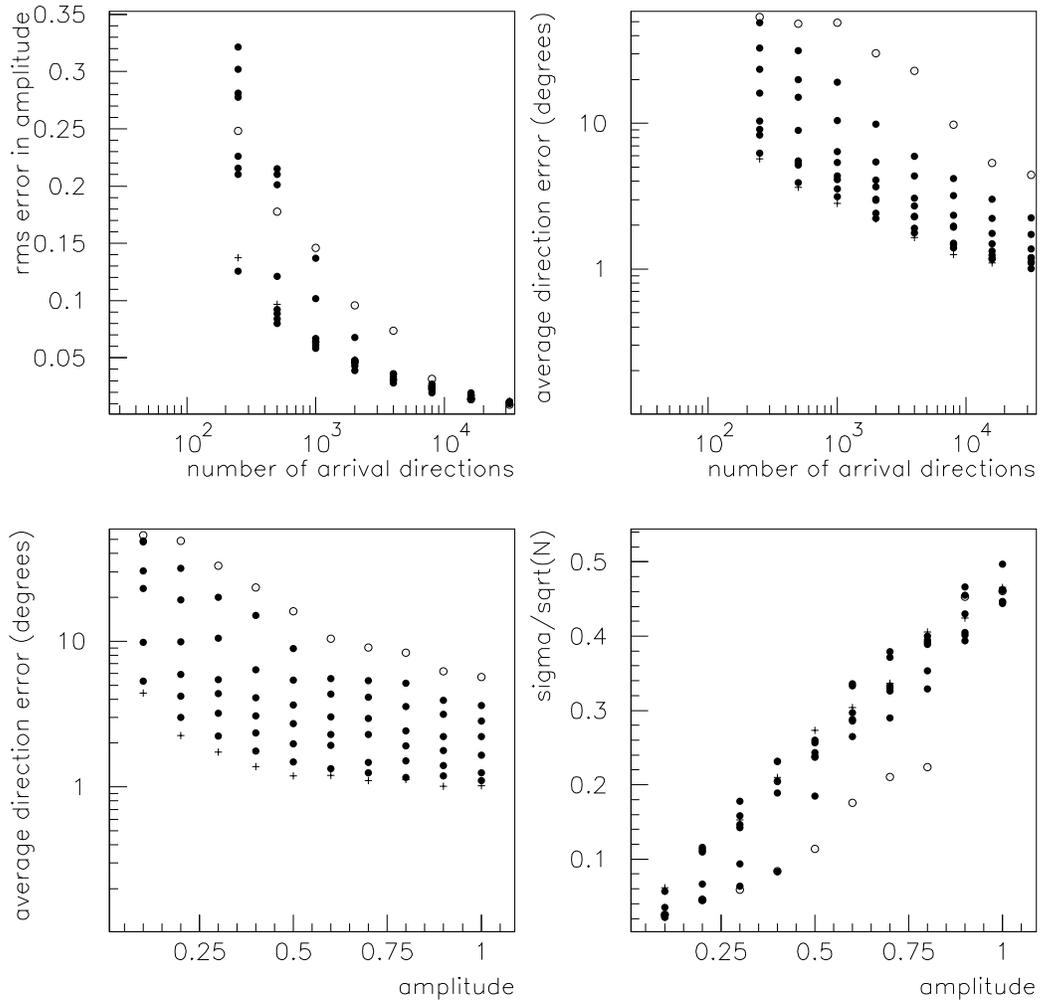,width=6.0in}
\vspace{1cm}
\caption{Similar to figure 2 for dipole sensitivity, the plots in
this figure indicate how sensitivity to {\it quadrupole}
anisotropy depends on the size of the data set $N$ and the
anisotropy amplitude $\alpha$.  
Symbol definitions are the same as in figure 2.  Multiplying the
ordinate of a point in the lower right plot by $\protect\sqrt{N}$
gives the number of sigmas deviation from isotropy.  Each point is
derived from an ensemble of 100 data sets.} 
\end{figure}

\section{Spherical harmonics}

For any data set of arrival directions (with full-sky exposure),
the anisotropy patterns can be fully characterized by the set of
spherical harmonic coefficients $a_{\ell m}$, in terms of which
the intensity function over the sphere is given by
$$I(\theta,\phi)=\sum^{\infty}_{l=1}\ \sum^{\ell}_{m=-\ell}\
a_{\ell m}\ Y_{\ell m}(\theta,\phi).$$

The coefficients $a_{\ell m}$ are given by
$$a_{\ell m} = \int\ I(\theta,\phi)\ Y_{\ell m}(\theta,\phi)\
d\Omega.$$ Real-valued spherical harmonics are used in this
paper, so the coefficients are real.  The real-valued $Y_{\ell m}$
functions are obtained from the complex ones by substituting
$$e^{im\phi} \longrightarrow \left\{ \begin{array}{ll}
\ \sqrt{2}\ sin(m\phi) & m<0 \\
\ 1 & m=0 \\
\ \sqrt{2}\ cos(m\phi) & m>0 \end{array} \right. $$

For a set of $N$ discrete arrival directions with non-uniform
relative expousre $\omega(\vec{u})$, the estimate for $a_{\ell m}$ is given by
$$a_{\ell m} = \frac{1}{\cal N} \sum_{i=1}^N\ \frac{1}{\omega_i}\
Y_{\ell m}(\vec{u}^{(i)})$$
where $\omega_i$ is the relative exposure at arrival direction
$\vec{u}^{(i)}$ and $\cal N$ is the simple sum of the weights
$\frac{1}{\omega_i}$.  

The upper left plot in figure 4 shows a scatter plot of 2921
directions to extragalactic infrared sources with $z<0.01$,
obtained from the NASA/IPAC Extragalactic Database (NED)
\cite{NED}.  There are certainly selection effects in these
directions, but they are used here only as an example of an
anisotropic celestial distribution.  The spherical harmonic
coefficients $a_{\ell m}$ for this distribution are plotted to
the right of that scatter plot.  There are 440 coefficients
plotted for $1\leq \ell \leq 20$.  Each set of $2\ell +1$
coefficients is plotted for each $\ell$ over an interval of 0.4
units on the abscissa.  These $a_{\ell m}$ constitute a
``fingerprint'' of the anisotropy.  They define a celestial
intensity function that is a smoothed version of the scatter
plot.  The prominent coefficients $a_{11}$ and $a_{22}$ in this
example result from the strong excess of the Virgo Cluster seen
in the left central part of the scatter plot.  Virgo is at
declination 12.7 degrees and right ascension 187 degrees, and the
$a_{\ell m}$ coefficients are derived here using that equatorial
coordinate system (not the plotted supergalactic coordinate
system).

\begin{figure}
\psfig{figure=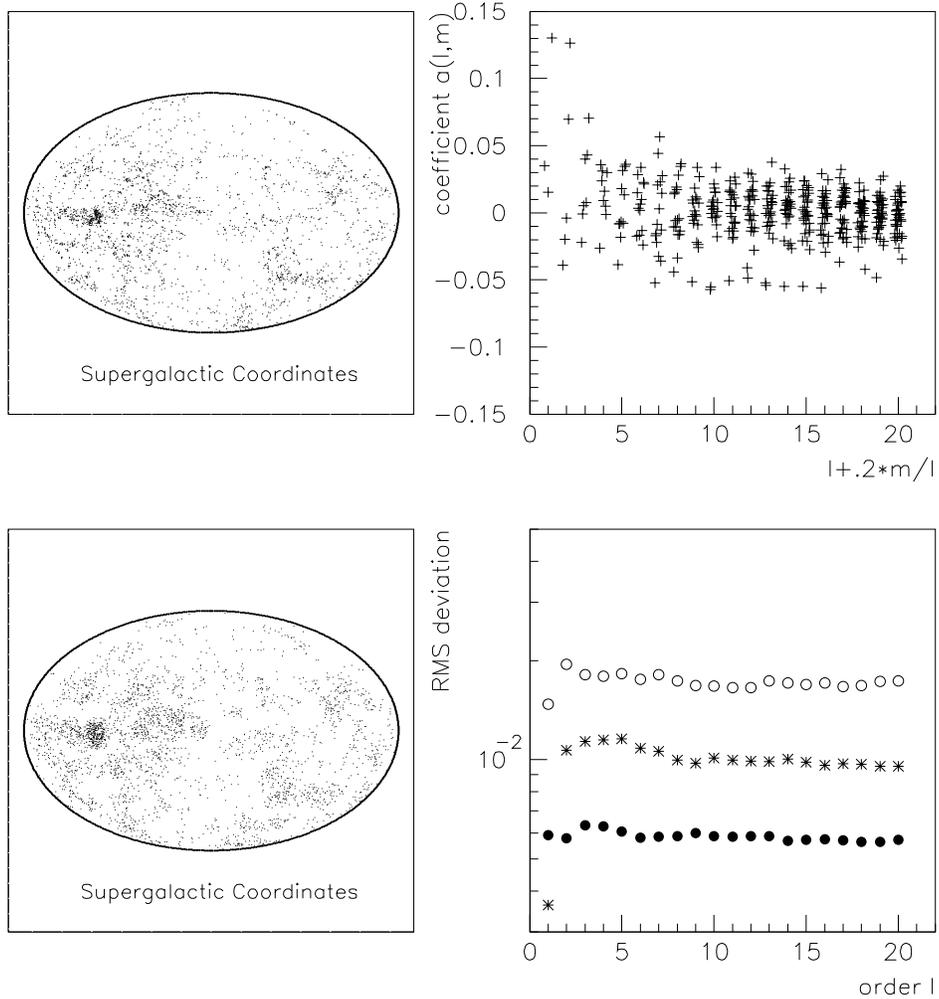,width=6.0in}
\vspace{1cm}
\caption{This figure indicates how anisotropy in arrival
directions is characterized by spherical harmonic coefficients
$a_{\ell m}$.  The upper left plot is 2921 directions to
extragalactic infrared
sources.  The upper right plot shows the spherical harmonic
coefficients out to $\ell=20$ (using the equatorial coordinate
basis).  The lower left scatter plot shows arrival directions
sampled from the smoothed celestial intensity function defined by
those $a_{\ell m}$ coefficients (with the Auger exposure also
imposed).  Plotted at the lower right are the RMS differences
between $a_{\ell m}$ coefficients derived from such sample
simulations and those shown at the upper right.  Open circles in
the lower right correspond to a simulation with 250 sampled
directions, asterisks to 1000 sampled directions, and the filled
circles correspond to the simulation shown in the lower left with
2921 sampled arrival directions.}
\end{figure}

To illustrate how well the $a_{\ell m}$ coefficients characterize
the anisotropy, arrival directions can be sampled from the
intensity function that they define.  The lower left plot is a
scatter plot with the same number of directions (2921) based on
the relative intensity function
$$I=\sum^{20}_{l=1}\sum^{\ell}_{m=-\ell}\ a_{\ell m}\ Y_{\ell
m}.$$
The Auger exposure function has also been imposed.  The lower
left plot should not be identical to the upper left plot because
of the Auger exposure simulation as well as the random sampling
from the smoothed celestial anisotropy function.  It clearly does
have the same primary features, however.  

The lower right plot in figure 4 summarizes how well the
anisotropy is determined this way as a function of the number of
arrival directions.  The filled circles in that plot correspond
to the displayed scatter plot with 2921 arrival directions.  The
$a_{\ell m}$ coefficients were derived from that simulation data
set (using the relative exposure weights) and compared with
those derived from the infrared source distribution.  For each
$\ell$, the RMS difference in the $a_{\ell m}$ values is plotted.
One can see that the typical error in $a_{\ell m}$ is small
compared to the significant coefficients in the upper right plot
that characterize the anisotropy.  The asterisks in the lower
right plot are derived in the same way using a simulation with
1000 arrival directions.  The open circles are the RMS coefficient
differences resulting from a simulation with just 250 sampled
arrival directions.  The anisotropy fingerprint in this example is still
measurable with 250 directions, although the RMS uncertainty in
the $a_{\ell m}$ coefficients grows as $1/\sqrt{N}$ as $N$ decreases.

\section{The angular power spectrum}

The angular power spectrum is the average $a_{\ell m}^2$ as a
function of $\ell$:
$$C(\ell) = \frac{1}{2l+1}\ \sum_{m=-\ell}^{\ell}\ a_{\ell
m}^2.$$   The power in mode $\ell$ is sensitive to variations
over angular scales near $1/\ell$ radians.  The angular power
spectrum provides a quick and sensitive method to test for
anisotropy and to determine its magnitude and characteristic
angular scale(s).

\begin{figure}
\psfig{figure=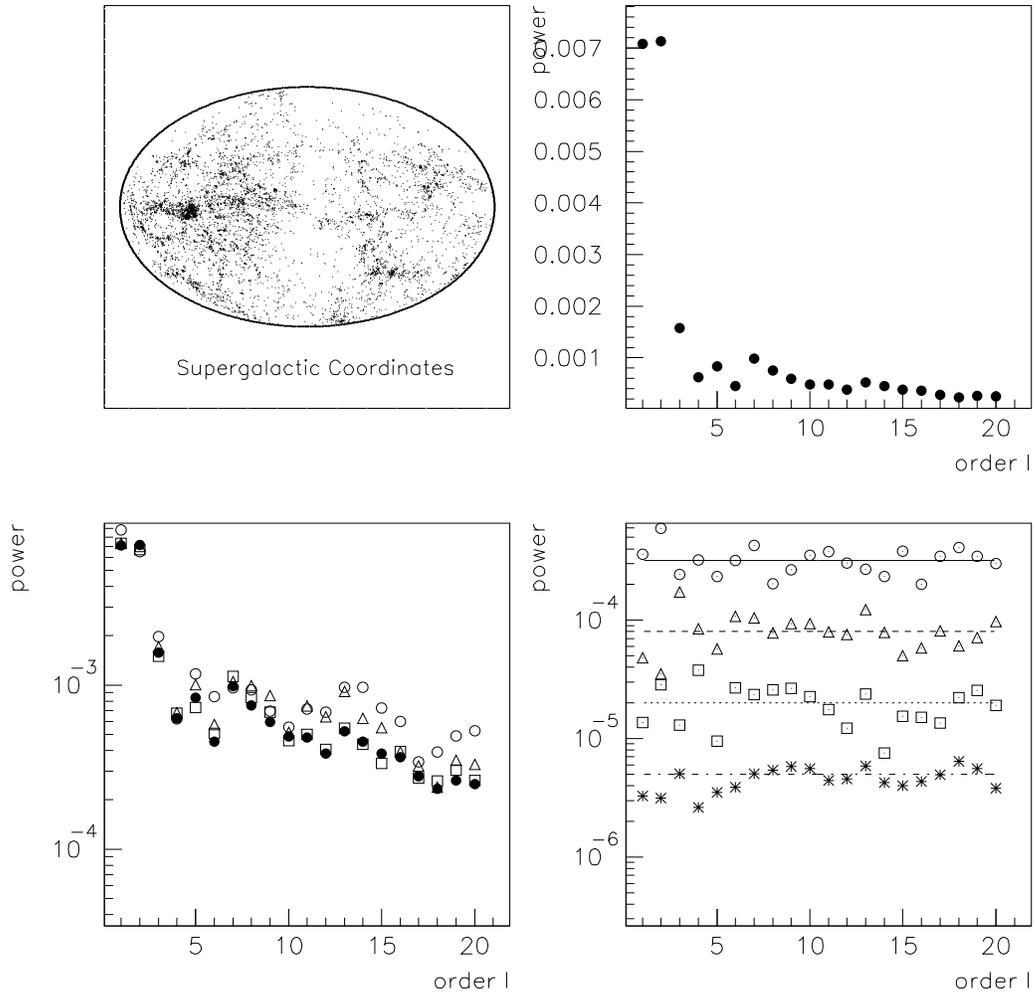,width=6.0in}
\vspace{1cm}
\caption{The upper left plot shows the celestial distribution of
7321 galaxy directions with $z<.01$ in supergalactic coordinates.
The upper right plot is the power spectrum obtained using uniform
exposure.  The lower left plot compares this same power spectrum
(filled circles) with power spectra obtained from simulation data
sets of non-uniform exposure and reduced numbers of directions.
Open circles result from a data set of 250 directions sampled
from the galaxy distribution with the Auger relative exposure.
The triangles pertain to a data set with 1000 directions, and
squares to 4000 directions.  The lower right plot shows the
expected power resulting from the same type of finite sampling
from an isotropic intensity.  The symbols are the same as in the
lower left plot, with asterisks representing a data set of 16,000
sampled directions.}
\end{figure}

As an example, consider the distribution of galaxies shown in the
upper left plot of figure 5.  These are all galaxies with
redshift $z<0.01$ (also obtained from NED), and they are plotted
in the supergalactic coordinate system.  The Virgo cluster is the
highest density region toward the left in this plot.  There are
7321 galaxy positions plotted.  The angular power spectrum (with
no exposure correction) is shown out to $\ell=20$ in the upper
right plot of that figure.  There is excess power at all
$\ell$-values, but especially for the dipole and quadrupole
moments ($\ell$ = 1, 2) due to the high intensity from Virgo and other
parts of the supergalactic plane.

The lower left plot in figure 5 indicates how sensitivity
to the power spectrum is affected by the number of arrival
directions (with non-uniform exposure as is expected for the Auger
Observatory).  Open circles in that plot are the power spectrum
derived using a data set of 250 arrival directions.  Those
directions were obtained by randomly sampling from the 7321
galaxy directions (without replacement) and rejecting a sampled
direction if a random number between 0 and 1 fell above the
relative exposure $\omega$ evaluated at that direction.  The 250
directions therefore represent a simulation Auger data set if
each galaxy direction were to have equal probability of being a
cosmic ray arrival direction.  The open circles in the lower left
plot are a decent approximation to the power spectrum (filled
circles) of the ``true'' power spectrum defined by all 7321
directions with uniform exposure.  The triangles in that plot
represent the power spectrum for 1000 arrival directions sampled
in the same way from the 7321, and the squares are obtained from
4000 sampled with the Auger exposure.  It is clear that the
approximation to the true power spectrum improves as the data
set gets richer, but the gross information is already present
with 250 arrival directions.

The lower right plot of figure 5 indicates how much power is
expected due to fluctuations when directions are sampled from an
isotropic intensity (and biased for the non-uniform expousre).
The power is the same for all $\ell$-values and decreases like
$1/N$ as the number of arrival directions increases.  The power
spectrum of the galaxy distribution is well above this noise
level for all $\ell$-values for the cases $N$ = 4000 or 16,000.
For 250 directions, only the first prominent harmonics in the
lower left plot are clearly above the noise level indicated in
the lower right plot.

Any class of candidate objects (e.g. active galaxies, or active
galaxies with giant radio hot spots) has a celestial distribution
that can be compared with a cosmic ray map when the whole sky has
been surveyed with adequate sensitivity.  Full information about
the celestial distribution is provided by the set of coefficients
$a_{\ell m}$.  They can be tabulated out to $\ell=20$ in a list
of 441 numbers (including the monopole).  The angular power
spectrum is a coordinate-independent gross summary of the
features present in the celestial distribution.  For example, you
may learn from it that there is a large quadrupole moment, but
you do not learn if the quadrupole has axial symmetry or the
orientations of its principal axes.  Full anisotropy information
is given by the 441 $a_{\ell m}$ coefficients, not the 20 $C(l)$
powers.

The magnitude of the angular power $C(l)$ for larger
$\ell$-values may contain useful information in the case that
cosmic rays come from a limited number of discrete sources. The
solid angle extent of the typical source affects the power at
large values of $\ell$.  Figure 6 displays an example in which
there are 50 sources of equal flux with positions sampled
randomly on the sky.  Three different sky plots are shown,
corresponding to different hypotheses about how much the arrival
directions are dispersed from the source direction.  In the upper
left plot, sampled arrival directions are accepted only if they
lie within $10^{\circ}$ of one of the sources.  In the upper
right plot, they are accepted only within 5 degrees of a source,
and only within 1.5 degrees in the lower left plot.  The graph in
the lower right shows the power spectra for the three different
simulations.  The power at low $\ell$-values is governed by the
chance pattern in the distribution of source positions.  For
$\ell \simgt 10$, however, the power clearly increases as the
amount of source smearing decreases.  The high end of the
measurable angular power spectrum is sensitive to anisotropy
structure on that finer scale.

\begin{figure}
\psfig{figure=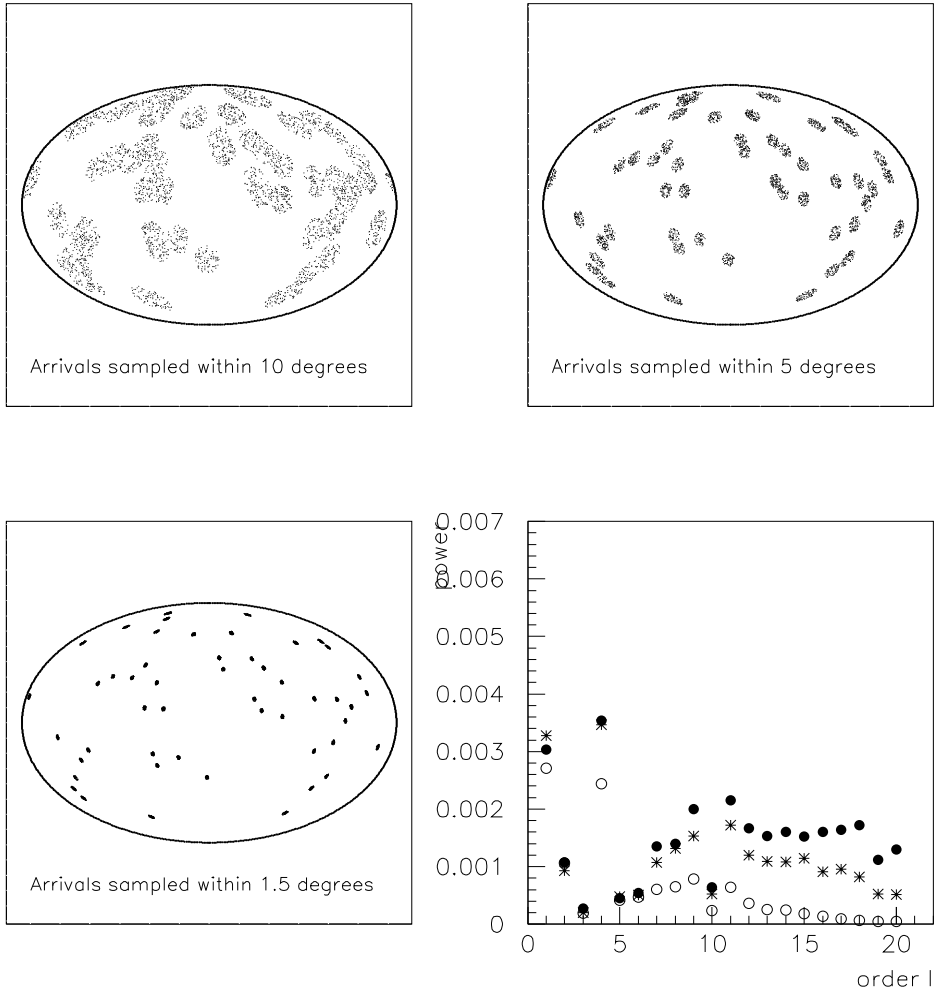,width=6.0in}
\vspace{1cm}
\caption{This study explores how the angular power spectrum from
a set of discrete sources should change with the amount of
smearing by magnetic dispersion.  Fifty hypothetical source
positions were sampled isotropically.  A set of 4000 arrival
directions were then sampled.  An arrival direction was accepted
if it were within an angular distance $\theta$ of any source (and
was subjected to possible exclusion in accordance with Auger
non-uniform exposure).  Results for three different values of
$\theta$ are shown: $\theta=10^{\circ}$ in the upper left,
$\theta=5^{\circ}$ in the upper right, and $\theta=1.5^{\circ}$
in the lower left.  Power spectra for the three cases are shown
in the graph at the lower right (open circles for
$\theta=10^{\circ}$, asterisks for $\theta=5^{\circ}$, and filled
circles for $\theta=1.5^{\circ}$.}

\end{figure}

\section{Discussion}

There is great advantage in a cosmic ray observatory having
exposure to the entire celestial sphere, especially if the
relative exposure is nearly uniform.  In that case, scatter plots
of arrival directions are immediately interpretable, and eyeball
evaluations can readily identify discrete sources or large-scale
patterns.  Discrete sources will be identified with equal
sensitivity anywhere in the sky.  If no such sources are found,
the flux upper limits will be uniform over the sky.  

At the highest energies there is no proven anisotropy.  Unlike
the COBE anisotropy analysis, it is not necessary to subtract a
large known dipole pattern and a myriad of uninteresting
foreground sources.  Any cosmic ray deviations from isotropy will
be of immediate interest.  The search for cosmic ray anisotropy
is more similar to the case of gamma-ray bursts, where
expectations and early claims of anisotropy were not supported by
additional data.

The role of an observatory is to map the sky and make the results
available to the scientific community.  This is highly
challenging for an observatory without full-sky coverage.
Measurements in that case are made with different sensitivity in
different parts of the sky, and nothing at all can be said about
a large hole where the exposure is zero.  Certainly it is not
possible to perform the full-sky integrations that are required
to measure the multipoles of the celestial cosmic ray
intensity.  In this paper, frequent use is made of the inverse of
the relative exposure, $1/\omega$.  Such methods obviously fail if
the relative exposure anywhere becomes infinitesimal or zero.

To underscore the difficulty of anisotropy analysis without
full-sky coverage, one can cite the work by Wdowczyk and
Wolfendale \cite{Wolfendale}.  In that paper, the authors argue
that the cosmic ray intensity measurements support a model of
excess arrivals from equatorial galactic latitudes.  The argument
is based on the same data that two experimental groups had
previously used in support of a gradient in galactic latitude
that suggested an excess from southern latitudes relative to
northern latitudes.  In effect, because those northern detectors
had poor exposure for southern galactic latitudes, Wdowczyk and
Wolfendale were able to argue that the data supported a
quadrupole distribution rather than a dipole distribution.
Neither a dipole moment nor a quadrupole moment can be measured
without full-sky coverage.

The techniques outlined in this paper pertain to any full-sky
detector.  Non-uniformity in celestial exposure is not hard to
handle, provided it is well determined and there is adequate
exposure to all parts of the sky.  The true cosmic ray intensity
is mapped with a sensitivity that depends primarily on the total
number of detected arrival directions.  This number $N$ is
related to particle energy and observing time for any detector of
known acceptance.  This relationship for the Auger Observatory
was given in the Introduction.

While complete information about anisotropy is encoded in the
$a_{\ell m}$ coefficients (tied to some specified coordinate
system), important gross properties of the anisotropy are
characterized by the (coordinate independent) angular power
spectrum $C(l)$.  One can tell, for example, if there is a strong
dipole or quadrupole moment.  Such large-scale patterns are
expected in many theories.  It should be noted, however, that
$C(1)$ gives the dipole moment but not its direction.  It is
obviously important whether the dipole points toward the galactic
center, toward Virgo, toward Cen A, or in some unexpected
direction. Similarly, all components of the quadrupole tensor are
of interest, not just the average of their squares, $C(2)$.
Nevertheless, the angular power spectrum provides a powerful tool
for discriminating between viable and non-viable theories without
detailed investigation.  Also, the higher-order moments of the
angular power spectrum can quantitatively characterize whatever
clumpiness may exist in a map of arrival directions.

The techniques and examples mentioned in this paper are only
representative of the powerful analysis methods that become
possible with full-sky observatories.  Data sets from such
observatories will open a rich field of anisotropy study.  The
primary goal seen from the present time is the discovery of the
highest energy cosmic ray origins.  If that objective is
accomplished (perhaps even before full-sky data sets are
available), then the observed patterns from a known source
distribution will be analyzed to infer properties of magnetic
fields in the galactic halo and in intergalactic space.

A full-sky observatory has the ability to summarize completely
the anisotropy information with the use of the spherical harmonic
expansion coefficients $a_{\ell m}$.  A table of coefficients
(perhaps with multiple columns for different energy cuts) will
provide the whole story.  As has been done in this paper, those
coefficients will be reliably corrected for the observatory's
non-uniform exposure.  Detailed anisotropy analysis will no
longer require privileged access to detector data.  The published
anisotropy fingerprint encoded in the $a_{\ell m}$ spherical
harmonic coefficients can be matched against any theoretical
suspect by any interested investigator.

\vspace{1.5cm}
\noindent {\it Acknowledgement.}  It is a pleasure to thank Brian
Fick for many helpful discussions about these matters.

\end{document}